\documentstyle{aipproc}

\begin{document}
\title{Hadron Spectroscopy from the Lattice}

\author{Chris Michael}
\address{Division of Theoretical Physics,
 Department of Mathematical Sciences, \\
University of Liverpool, Liverpool L69 3BX, UK.\\}

\maketitle

\begin{abstract}

 Lattice QCD determinations appropriate to hadron spectroscopy are
reviewed with emphasis on the  glueball and hybrid meson states in the
quenched approximation. Hybrids are  discussed for heavy and for light
quarks. The effects of sea quarks (unquenching) are  explored.
Heavy-light systems are presented - particularly excited B mesons.

\end{abstract}

\section*{Introduction}

   Quantum Chromodynamics is generally acknowledged to be the theory of
hadronic  interactions. Its perturbative features are well understood
and provided the  main motivation for its adoption. To give a complete
description of hadronic  physics, it is essential to develop
non-perturbative methods to handle QCD. QCD is a quantum field theory
which  needs to be regulated in order to have a well defined
mathematical approach.  Inevitably such a regularisation will destroy
some of the symmetries of QCD.  For example  dimensional regularisation
is often used in perturbative  studies and it breaks the 4-dimensional
Lorentz invariance. Similarly lattice regularisation as proposed by
Wilson~\cite{wilson} breaks Lorentz invariance since a hypercubic
lattice of space-time points is invoked. The key feature of Wilson's
proposal  is that gauge invariance is exactly retained. Then the
approach to the continuum limit (as the lattice spacing $a$ is reduced
to zero) can be shown to be  well defined. Using Monte Carlo methods to 
explore lattice QCD, reliable predictions can now be made for continuum 
quantities as I shall discuss. Moreover, in the continuum limit, the Lorentz 
invariance is found to be fully restored.

  Lattice QCD needs as input the quark masses and an overall scale
(conventionally  given by $\Lambda_{QCD}$). Then any Green function can
be evaluated by taking an average of suitable combinations of the
lattice fields in the vacuum samples. This allows masses to be studied 
easily and matrix elements (particularly those of weak or
electromagnetic currents)  can be extracted straightforwardly.
Scattering and hadronic decays are only  accessible in a rather limited
way. 

  Unlike experiment, lattice QCD can vary the quark masses and can also 
explore different boundary conditions and sources. This allows a wide range of 
studies which can be used to diagnose the health of phenomenological models
as well as casting light on experimental data.

 One very special case is of considerable interest: this is {\em
quenched} QCD where the sea-quark masses are taken as infinite. This
suppresses  quark loops in the vacuum completely, leaving just the full
non-perturbative  gluonic interactions. This gluonic vacuum turns out to
reproduce most of  the salient features of QCD. It is also a very
convenient approximation to  use for comparison with phenomenological
models. Quenched QCD is computationally rather easy to study and the
precise results allow the continuum limit to be extracted  reliably. 
Studies with sea quark effects included (known as dynamical fermion 
studies) are computationally much more demanding. I discuss the current
situation  in this area in the last section.

\section*{Glueballs}

Glueballs are defined to be hadronic states made primarily from gluons.
The full non-perturbative gluonic interaction is included in quenched
QCD.  In the quenched approximation, there is no mixing between such
glueballs  and quark - antiquark mesons. A study of the glueball
spectrum in quenched QCD  is thus of great value. This will allow
experimental searches to be  guided as well as providing calibration for
models of glueballs.

In principle, lattice QCD can study the meson spectrum as the sea quark
mass  is decreased towards experimental values. This will allow the
unambiguous glueball states in the quenched approximation to be tracked
as the sea quark effects are increased.  It may indeed turn out that no
meson in the physical spectrum is primarily a glueball - all states are 
mixtures of glue,  $q \bar{q}$, $q \bar{q} q \bar{q}$, etc.  Studies 
conducted so far show no significant change of the glueball spectrum as
dynamical quark effects are added - but  the sea quark masses used are
still rather large~\cite{sesam}.

In lattice studies, dimensionless ratios of  quantities are obtained. To
explore the glueball masses, it is appropriate to combine  them with
another very accurately measured quantity to have a dimensionless 
observable. Since the potential between static quarks is very accurately
measured from the lattice (see the next section for more details), it is
 now conventional to use $r_0$ for this comparison.  Here $r_0$ is
implicitly defined by $r^2 dV(r)/dr = 1.65$ at $r=r_0$. In practice
$r_0$ may be related to the string tension  $\sigma$ by $r_0
\sqrt{\sigma}=1.18$. 

 Theoretical analysis  indicates that for the quenched approximation 
the dimensionless ratio $mr_0$ will differ from the continuum  limit
value by corrections of order $a^2$.  Thus in Fig.~1 the masses are
plotted versus the lattice spacing $a^2$ for the $J^{PC}$=$0^{++}$ and
$2^{++}$ glueballs. The straight lines then show the continuum limit
obtained  by extrapolating to $a=0$. As can be seen, there is
essentially no need for data  at even smaller $a$-values to further fix
the continuum value. The values shown  correspond to
$m(0^{++})r_0=4.33(5)$ and $m(2^{++})r_0=6.0(6)$.  Since several
lattice groups~\cite{DForc,MTgl,ukqcd,gf11} have measured these 
quantities, it is reassuring to see that the purely lattice observables
are in  excellent agreement. The publicised difference of quoted
$m(0^{++})$ from  UKQCD~\cite{ukqcd} and GF11~\cite{gf11} comes entirely
from relating quenched lattice  measurements to values in GeV as I now
discuss.

\begin{figure}[bt] 
\vspace{11.1cm} 
\includegraphics{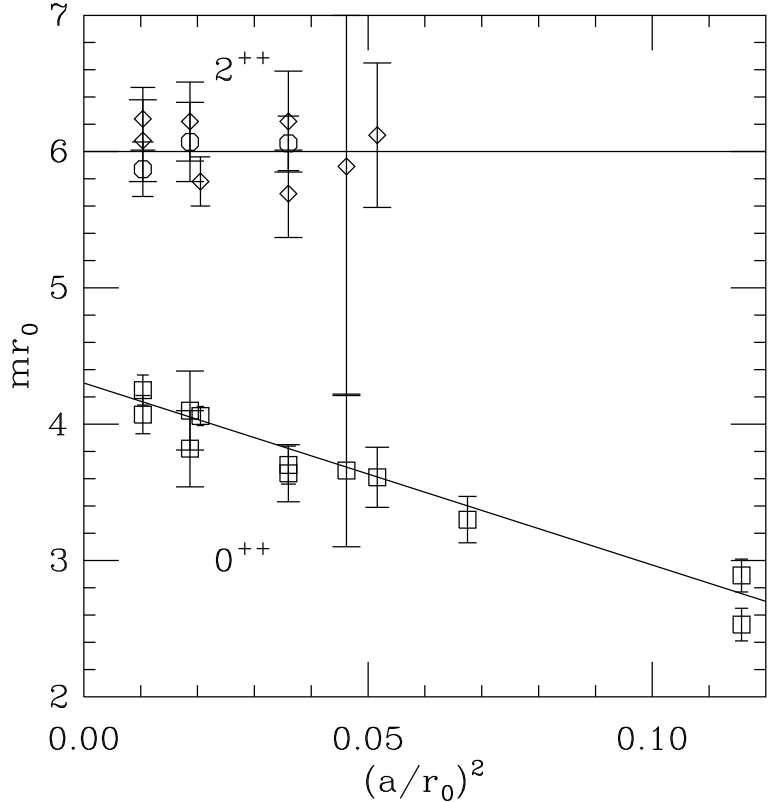}
 \caption{ The value of mass of the  $J^{PC}=0^{++}$ and $2^{++}$
glueball states from refs{\protect\cite{DForc,MTgl,ukqcd,gf11}} in units
of $r_0$ where $r_0 \approx 0.5$ fm. The $T_2$ and $E$ representations
are shown  by  octagons  and    diamonds respectively and their agreement
indicates the restoration of rotational invariance for the $2^{++}$
state.  The straight lines  show  fits describing the  approach to the
continuum limit as $a \to 0$.
   }
\end{figure}

In the quenched approximation, different hadronic observables differ
from experiment  by factors of up to 10\%. Thus using one quantity or
another to set the scale, gives an overall systematic error.  Here I
choose to set the scale by taking the conventional value of the string
tension, $\sqrt{\sigma}=0.44$GeV, which then corresponds to
$r_0^{-1}=373$ MeV. An overall systematic error of 10\% is then to be
included to any  extracted mass. This yields $m(0^{++})=1611(30)(160)$
MeV and  $m(2^{++})=2232(220)(220)$ MeV where the second error is the
systematic  scale error. Note that these are glueball masses in the
quenched approximation -  in the real world significant mixing with $q
\bar{q}$ states could modify these values substantially.

Recently a lattice approach using a large spatial lattice spacing with
an  improved action and a small time spacing has been used to study
glueball  masses. The results~\cite{mpglue} are that
$r_0m(0^{++})=3.98(15)$,  $r_0m(2^{++})=5.85(2)$, $r_0m(1^{+-})=7.21(2)$
and $r_0 m'(2^{++})=8.11(4)$. There remains a small discrepancy with the
result for the $0^{++}$ glueball obtained above (4.33(5)) from lattice
spacings much closer to the continuum limit. When this is fully 
understood, the new method looks to be very promising for  access to
excited glueball masses.

I have focussed on the scalar and tensor glueball results because these
are the  lightest and best measured states in lattice studies. The
glueball spectrum has been  extracted for all $J^{PC}$
values~\cite{MTgl,ukqcd}. Results are shown in Fig.~2.  One signal of
great interest would be  a glueball with $J^{PC}$ not allowed for $q
\bar{q}$ - a spin-exotic glueball or {\em oddball}. These states are
shown  in Fig.~2 to be high lying: at least above $2m(0^{++})$. Thus
they are  likely to be in a region very difficult to access
unambiguously by experiment.

\begin{figure}[bt]
\vspace{10.3cm} 
\includegraphics{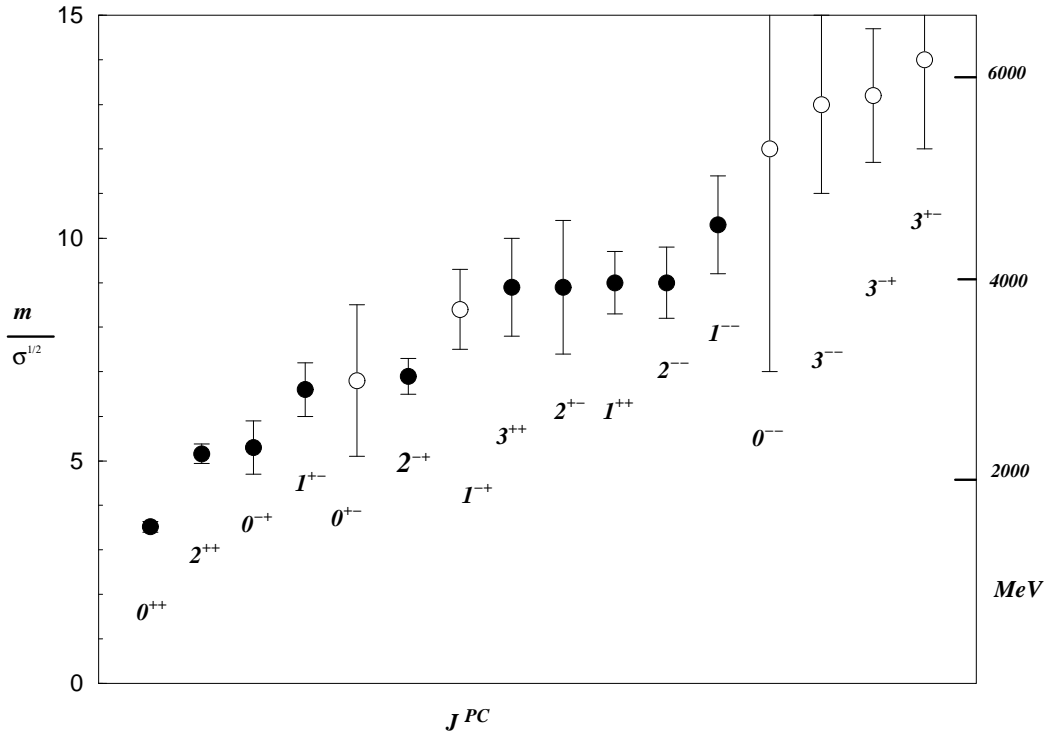}
 \caption{ The  mass of the  glueball states  with quantum  numbers
$J^{PC}$ from ref{\protect\cite{ukqcd}}.  The scale is set by
${\protect\sqrt{\sigma}} \approx 440 $ MeV which yields the right hand
scale in MeV. The solid circles represent mass determinations whereas the
open circles  are upper limits.
   }
\end{figure}

 The only other candidate for a relatively light glueball is the
pseudoscalar. Values quoted of $r_0m(0^{-+})= 5.6(6),\ 7.1(1.1)$ and
5.3(6) from refs\cite{MTgl,ukqcd} suggest  an average of 6.0(1.0), not
appreciably lighter  than the tensor glueball. This is confirmed by
preliminary results from the group  of ref\cite{mpglue} that the
pseudoscalar is heavier than the tensor glueball.

Within the quenched approximation, even though the $0^{++}$ glueball
width is zero, it is possible in principle to measure the  decay matrix
element between a $0^{++}$ glueball and two pseudoscalar $q\bar{q}$ 
mesons. This then allows an estimate of the width which would arise 
when going beyond the quenched approximation.  A first attempt to study
this~\cite{gdecay} yields estimated widths of order 100 MeV. Even though
this lattice calculation is very exploratory, it does indicate that very
wide widths to two pseudoscalars are not expected. A more realistic
study  would involve mixing with the $q \bar{q}$ and $s \bar{s}$ scalar
mesons as  well as direct decays.

\section*{Heavy Quark Interactions }

In the limit  $m_Q \to \infty$, the heavy quark effective theory
describes a universal behaviour. For finite $m_Q$, corrections of order
$1/m_Q$  are expected. The simplest way to study the heavy quark limit
on a lattice is to  use static quarks. The potential energy $V(R)$
between a static quark and antiquark  at separation $R$ is readily
obtained. Then for heavy quarks, one may solve for  the spectra in this
potential using the Schr\"odinger equation in the  adiabatic
approximation.  The quenched lattice potential is well measured and is
found to have a form parametrised by 
 \begin{equation}
 V(R) = V_0 - {e \over R} + \sigma R
 \end{equation}  
 where $e$ is the coefficient of the Coulomb term and $\sigma$ is the
string tension. This expression shows that the potential  continues to
increase as $R$ is increased - this is confinement.

 A comparison from ref\cite{pm} of the spectrum in the quenched lattice
potential with  the $\Upsilon$ states is shown in  Fig.~3. The lattice
result is qualitatively similar to the experimental $\Upsilon$ spectrum.
The main difference is that the Coulombic part ($e$) is effectively too
small (0.28 rather than 0.48). This produces~\cite{pm} a ratio of   mass
differences $(1P-1S)/(2S-1S)$ of 0.71 to be compared  with the
experimental ratio of 0.78.  This difference in $e$ is understandable as
a consequence of the Coulombic force at short distances which would be
increased by $33/(33-2N_f)$ in perturbation theory in full QCD compared
to quenched QCD. We will return to discuss this.

\begin{figure}[bt!] 
\vspace{11cm} 
\includegraphics{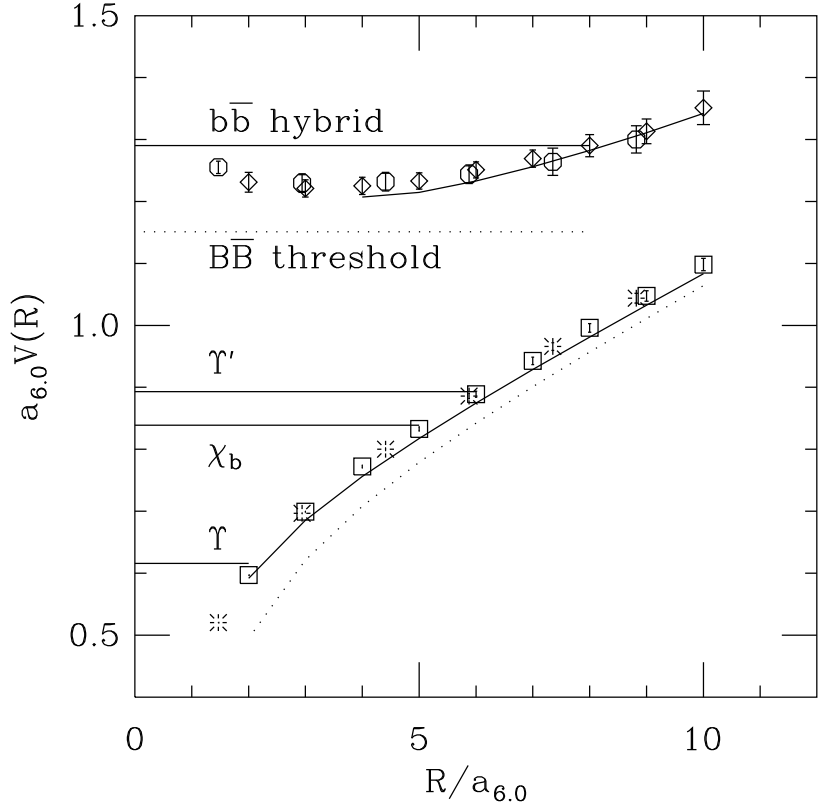}
 \caption{ Potentials $V(R)$ between static quarks  at separation $R$
for the  ground state (square and {\tt *}) and for the $E_u$ symmetry
which corresponds to the first excited state of the gluonic flux
(octagons and diamonds). Results in lattice units ($a_{6.0}^{-1}=2.02$
GeV) from  the quenched calculations of ref{\protect\cite{pm}} are shown
by symbols corresponding to different lattice spacings. For the ground
state  potential the continuous curve is an interpolation of the lattice
data while the dotted  curve with enhanced Coulomb term fits the
spectrum and yields the masses shown. The lightest hybrid level in the
excited gluonic potential is  also shown.
   }
\end{figure}

The situation of a static quark and antiquark is a very clear case in
which to discuss hybrid  mesons which  have excited gluonic
contributions. A discussion of the colour  representation of the quark
and antiquark is not useful since they are at different space positions
and the combined colour is not gauge invariant. A better criterion is 
to focus on the spatial symmetry of the gluonic flux. As well as the
symmetric ground state of the colour flux between two  static quarks,
there will be excited states with different symmetries.  These were
studied on a lattice~\cite{liv} and the conclusion was that the $E_u$
symmetry (corresponding to flux  states from an  operator which is the
difference of U-shaped paths from quark to antiquark of the form $\,
\sqcap - \sqcup$) was the lowest lying gluonic excitation.  Results  for
this potential are shown in Fig.~3.

This gluonic excitation corresponds to a component of angular momentum
of one unit along the quark antiquark axis. Then one can solve for the
spectrum of hybrid mesons using the Schr\"odinger equation in the
adiabatic approximation. The spatial wave function necessarily has non
zero angular momentum and  the lightest states correspond to
$L^{PC}=1^{+-}$ and $1^{-+}$. Combining with the  quark and antiquark
spins then yields~\cite{liv} a set of 8 degenerate hybrid states with
$J^{PC}=1^{--},\ 0^{-+},\ 1^{-+},\ 2^{-+}$ and    $1^{++},\ 0^{+-},\
1^{+-},\ 2^{+-}$  respectively. These contain the  spin-exotic states
with $J^{PC}=  1^{-+},\ 0^{+-}$ and $2^{+-}$ which will be of special
interest. 

 Since the lattice calculation of the ground state and hybrid masses
allows  a direct prediction for their difference, the result for this
8-fold degenerate hybrid level is illustrated in Fig.~3  and
corresponds~\cite{pm} to masses of 10.81(25) GeV for $b\bar{b}$ and 
4.19(15) GeV for $c \bar{c}$. Here the errors take into account the
uncertainty in setting the ground state mass using the quenched
potential as discussed above. Recently a different lattice
technique~\cite{morn} has been used to  explore the excited gluonic
levels in the quenched approximation. The results above are confirmed
and preliminary values quoted for the lightest hybrid mesons are 10.83
and 4.25 GeV respectively for $b\bar{b}$ and $c \bar{c}$ with no error
estimates given.

 The quenched lattice results, after adjusting to take account  of the
measured $\bar{b} b $ spectrum,  suggest that the lightest hybrid mesons
lie  above the open $B \bar{B}$ threshold by about 270 MeV.  This can
also be studied by  comparing directly the lattice hybrid masses with
twice the quenched lattice  masses for the $B$ meson~\cite{sommer}.
Using quenched  results from the smallest lattice spacing ($\beta=6.2$)
available with clover-improved fermions~\cite{ewing} yields $E_{\rm
hybrid} - 2 m_{B} \approx  140(80)$ MeV. This estimate is somewhat
smaller than that  obtained above.  In both cases, however, the hybrid 
levels lie above the open threshold  and are likely to be relatively
wide  resonances. Another consequence is that the very flat potential
implies a very extended wavefunction: this has the implication that the
wavefunction at the origin will be small,  so hybrid vector states will
be weakly produced from $ e^+ e^-$.

 It would be useful to explore the splitting among the 8 degenerate
$J^{PC}$ values obtained. This could come from different excitation 
energies in the $L^{PC}=1^{+-}$ (magnetic) and $1^{-+}$
(pseudo-electric) gluonic excitations, spin-orbit terms, as well as
mixing between hybrid states and $Q\bar{Q}$ mesons with non-exotic spin.
One way to study this on a lattice is to use the  NRQCD formulation
which describes non-static heavy quarks which propagate 
non-relativistically. Preliminary results for hybrid excitations from
several  groups~\cite{manke,collins,mornnrqcd} give at present similar
results to those with the static approximation as described above, but
with  some additional evidence, namely that the magnetic excitations
(which include the $1^{-+}$ spin  exotic) are lighter than the
pseudo-electric ones.  Future NRQCD results may be more precise and able
to establish the splittings among different states.


\section*{Light Quark Interactions }

 Unlike very heavy quarks, light quark propagation in the gluonic vacuum
sample is very computationally intensive - involving inversion of huge
($10^7 \times 10^7$) sparse matrices. Current computer power is 
sufficient to study light quark physics thoroughly in the quenched 
approximation. The state of the art~\cite{yoshie} is the Japanese CP-PACS
Collaboration  who are able to study a range of large lattices (up to
about $64^4$) with a range of light quark masses. Qualitatively the 
meson and baryon spectrum of states made of  light and strange quarks is 
reproduced with discrepancies of order 10\% in the quenched approximation.

\begin{figure}[bt!] 
\vspace{11cm} 
\includegraphics{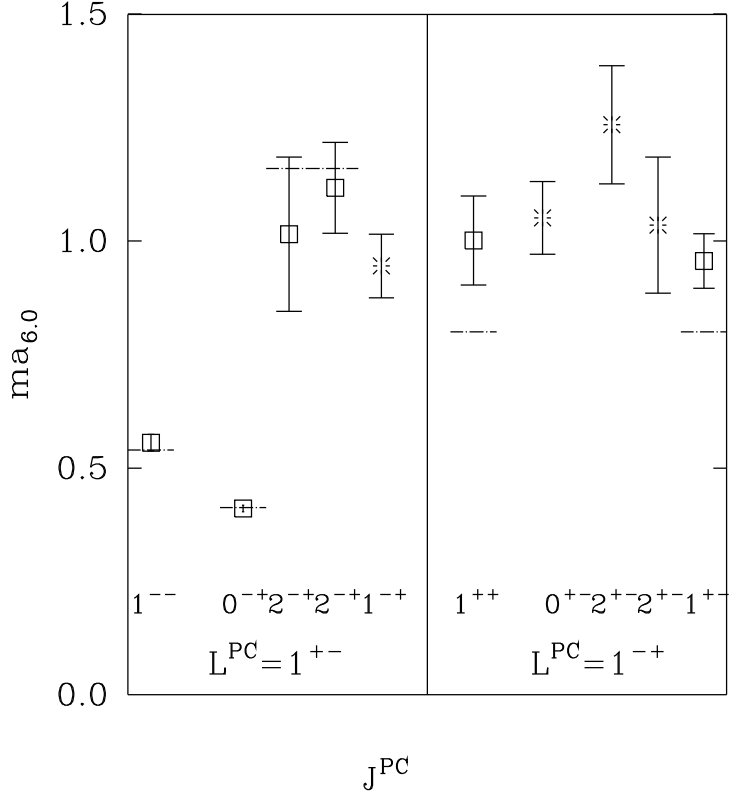}
 \caption{ The masses in lattice units (with $a_{6.0}^{-1} \approx 2$
GeV) of states of $J^{PC}$ built from hybrid operators with strange
quarks, spin-exotic ({\tt *}) and non-exotic (squares). The dot-dashed
lines are the mass values found for  $s\bar{s}$ operators. Results from 
 ref{\protect\cite{hybrid}}. 
   }
\end{figure}

 Here I will focus on hybrid mesons made from light quarks.  There will
be no mixing with $q \bar{q}$ mesons for  spin-exotic hybrid mesons  and
these are of special interest. The first study of this area was by the 
UKQCD Collaboration~\cite{hybrid} who used operators motivated by the 
heavy quark studies referred to above. Using non-local operators, they
studied  all 8 $J^{PC}$ values coming from $L^{PC}=1^{+-}$ and $1^{-+}$
excitations. The  resulting mass spectrum is shown in Fig.~4 where the
$J^{PC}=1^{-+}$ state  is seen to be the lightest spin-exotic state with
a statistical significance of 1 standard deviation. The statistical
error on the mass of this lightest spin-exotic meson  is 7\% but, to
take account of systematic errors from the lattice determination, a 
mass of 2000(200) MeV is quoted for this hybrid meson with $s \bar{s}$
light quarks. Although not  directly measured, the corresponding light
quark hybrid meson would be expected to be around 120 MeV lighter. In
view of the   small statistical error, it seems unlikely that the
$1^{-+}$ meson in the quenched approximation could lie as light as 1.4
GeV where there are experimental  indications for such a
state~\cite{bnl}. Beyond the quenched  approximation, there will be
mixing between such a hybrid meson and $q \bar{q} q \bar{q}$ states such
as $\eta \pi$ and this may be significant in the experimental 
situation.

One feature clearly seen in Fig.~4 is that non spin-exotic mesons
created  by hybrid meson operators have  masses  which are very similar
to those found when the states are created by $q \bar{q}$ operators.
This suggests that there is  quite strong coupling between hybrid and $q
\bar{q}$ mesons even in the quenched approximation. This would imply that 
the $\pi(1800)$ is unlikely to be a pure hybrid, for example.

A second lattice group has also evaluated hybrid meson spectra from
light quarks from queched lattices. They obtain~\cite{milc} masses of
the $1^{-+}$ state with statistical and various systematic errors of 
1970(90)(300) MeV, 2170(80)(100)(100) MeV and 4390(80)(200) MeV for $n
\bar{n}$,  $s \bar{s}$ and $c \bar{c}$ quarks respectively. For the 
$0^{+-}$ spin-exotic state they have a noisier signal but evidence that
it is heavier. They also explore mixing matrix elements between
spin-exotic hybrid  states and 4 quark operators.

\section*{Heavy-Light  Interactions }
 
 Here we discus the properties of $Q\bar{q}$ mesons and $Qqq$ baryons.
The  heavy quark effective theory indicates that many properties are
independent  of $m_Q$ to order $1/m_Q$. Many examples of this have been
studied using  lattice techniques: the Isgur-Wise function, the
pseudoscalar coupling  $f_Q \sqrt{m_Q}$ which yields $f_B$, the matrix
element for $B \bar{B}$ mixing $B_B$, and meson and baryon spectra and
decays.  Here I focus on one aspect which is rather easy to describe and which 
has recently been studied: the spectrum of excited states of the B meson.

  In the heavy quark limit, this $Q\bar{q}$ meson will be the `hydrogen 
atom' of QCD. Since the meson is made from non-identical quarks, charge
conjugation  is not a good quantum number.  States can be  labelled by 
$L_{\pm}$ where the coupling of the light quark spin to the orbital
angular momentum  gives  $j=L\pm {1 \over 2}$. In the heavy quark  limit
these states will be doubly degenerate since the heavy quark spin
interaction can be  neglected, so  the  $P_{-}$ state will have 
$J^{P}=0^+,\ 1^+$ while $P_{+}$ has $J^{P}=1^+,\ 2^+$, etc. A recent 
study~\cite{stoch} of this spectrum for $m_Q \to \infty$ gives the
preliminary results shown  in Fig.~5 for strange light quarks. Note that
most of these states have not been experimentally established for
excited $B$ mesons yet. A further  study of systematic errors and an
extrapolation to light quarks is in  progress. It will be interesting to
confirm the ordering of the $P_-$ and $P_+$ levels because it is
possible  that the long-range spin-orbit interaction could invert them,
making the $P_+$ lighter.

\begin{figure}[t!]  
\vspace{10cm} 
\includegraphics{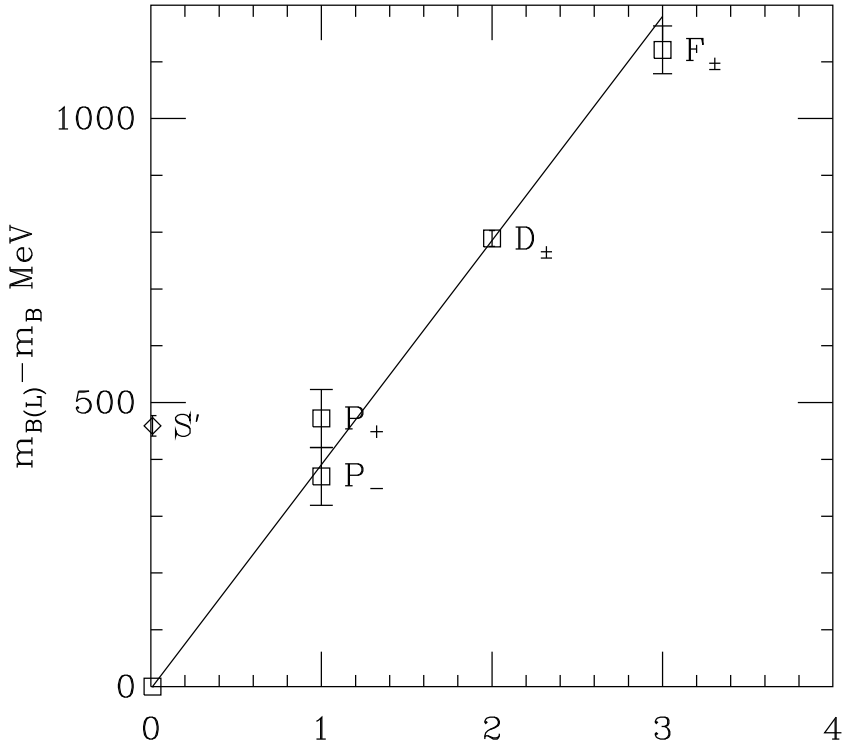}
 \caption{ The masses of excited $Q\bar{s}$ mesons from 
ref{\protect\cite{stoch}} versus $L$. The results are in the limit $m_Q
\to \infty$ and for the $L=2,\ 3$ states only the centre of gravity of
the two levels  was measured. The straight line is to guide the eye.
   }
\end{figure}

Another lattice approach, NRQCD, has also been used to study this  area
and has reported~\cite{alik} preliminary mass values for $b\bar{q}$  S
and P-wave mesons  and for $bqq$ baryons with light and strange quarks
in qualitative agreement with experiment.  This group has also given 
predictions for the $b \bar{c}$ meson~\cite{bc}.

For bound states of charm quarks, results for  mesons~\cite{boyle}
and baryons~\cite{stella} have been reported. These quenched approximation 
studies give a reasonable description of known states and several 
predictions for new ones.

\section*{Towards Full QCD}

 Algorithms exist which allow lattice simulation of full QCD with  sea
quarks of mass $m_{sea}$. This study needs lots of CPU power  since the
sea quark loops in the vacuum are represented effectively as  a long
range interaction between the gluonic degrees of freedom. Most  studies
to date have been exploratory with sea quark masses above the  strange
quark mass. In this regime, very little change from the quenched
approximation is seen in  physical predictions from the lattice. 

One area where specific changes are  expected is for the potential
between heavy quarks. At small separation $R$,  the Coulombic term is
expected to increase in strength and indeed some sign of this  has been
reported~\cite{sesamv}. At larger $R$,  signs of string breaking are
expected since a light quark antiquark pair can be produced  from the
vacuum to yield two $Q \bar{q}$ mesons with energy independent of $R$ 
at large separation. This has been explored by the SESAM and UKQCD
collaborations  and little sign of the effect is
seen~\cite{ukqcddf,sesamv}. 

Since it has been very difficult to see unambiguous signs of sea  quark
effects in the spectrum, it is possible that such effects turn on
non-linearly  as the sea quark mass is reduced. As an example, in
current studies the $\pi + \pi$ P-wave is heavier than the $\rho$ meson
so the $\rho$  cannot  decay. Further work is needed to reduce the sea
quark mass and to increase  the lattice size. Dedicated computing  power
of several hundreds of Gflops is available to lattice collaborations 
and progress in this area should now be possible.

\section*{Summary}

 Quenched lattice QCD is well understood and accurate predictions in the
continuum limit  are increasingly becoming available. Glueball masses of
$m(0^{++})=1611(30)(160)$ MeV;  $m(2^{++})=2232(220)(220)$ MeV and
$m(0^{-+})=2232(370)(220)$ MeV are  predicted where the second error is
an overall scale error. The quenched approximation  also gives
information on quark-antiquark scalar mesons and on hadronic decay 
matrix elements of glueballs. This mixing with $q \bar{q}$ mesons may
well result in  no clear experimental glueball candidate.  

 For hybrid mesons, there will be no mixing with $q \bar{q}$ for 
spin-exotic states and these are the most useful predictions. The
$J^{PC}=1^{-+}$ state is expected at 10.81(25) GeV for $b$ quarks; 
4.19(15) GeV for $c$ quarks, 2.0(2) GeV for $s$ quarks and 1.9(2) GeV 
for $u,\ d$ quarks. Mixing of spin-exotic hybrids with
$q\bar{q}q\bar{q}$ or equivalently with meson-meson  is  allowed and
will modify the  predictions from the quenched approximation.

 The lattice has proved a valuable source of information on the heavy
quark  effective theory. It gives information on the link between a $b$
quark  which enters the standard model and the $B$ meson of experiment.
Matrix elements  such as $f_B$ and $B_B$ have been measured. The spectra
of $b\bar{q}$ and $bqq$ hadrons have been predicted. 

 Much activity is currently underway to explore the effects of sea
quarks  of ever decreasing mass. Future teraflops computing facilities
will  be essential to obtain quantitative results both for hadronic
spectroscopy and  for the study of quark-gluon plasma at non-zero
temperature.

\end{document}